\documentclass[preprint,10pt,nofootinbib]{revtex4-1}
\usepackage[paper=letterpaper,margin=1.5in]{geometry}
\usepackage{amsmath,amssymb,amsfonts}
\usepackage{pstricks}
\usepackage{setspace}
\usepackage[varg]{txfonts}
\usepackage[]{graphicx}\graphicspath{{../graph/}%
{../FEL/"FEL at saturation (with Alberto)"/graph/}}
\usepackage{epstopdf}
\usepackage{hyperref}

\hypersetup{
  colorlinks,
  citecolor=green,
  linkcolor=blue}

\newcommand{\p}{\partial}
\newcommand{\const}{{\rm const}}

\renewcommand{\vec}[1]{\textnormal{\boldmath$#1$}}

\begin{document}

\title{A novel fast simulation technique for axisymmetric PWFA configurations in the blowout regime}

\author{P. Baxevanis and G. Stupakov}
\affiliation{SLAC National Accelerator Laboratory, Menlo Park, CA 94025}

\begin{abstract}

In the blowout regime of plasma wakefield acceleration (PWFA), which is the most relevant configuration for current and future applications and experiments, the plasma flow that is excited by the ultra-relativistic drive beam is highly nonlinear. Thus, fast and accurate simulations codes are indispensable tools in the study of this extremely important problem. We have developed a novel algorithm that deals with the propagation of axisymmetric bunches of otherwise arbitrary profile through a cold plasma of uniform density. In contrast to the existing PWFA simulation tools, our code PLEBS (PLasma-Electron Beam Simulations) uses a new computational scheme which ensures that the transverse and longitudinal directions are completely decoupled---a feature which significantly enhances the speed and robustness of the new method. Our numerical results are benchmarked against the QuickPic code and excellent agreement is established between the two approaches. Moreover, our new technique provides a very convenient framework for studying issues such as beam loading and short-range wakefields within the plasma cavity.

\end{abstract}

\maketitle

%
\section{Introduction}\label{sec:1}
%

The technique of plasma wakefield acceleration (PWFA), in which an intense, ultra-relativistic drive beam excites strong accelerating fields as it moves through a dense plasma column, is one of the most promising schemes for achieving the unprecedented acceleration gradients necessary for linear collider or compact free-electron laser (FEL) applications~\cite{Blumenfeld:2007qf,Litos:2014bh}. Unlike the early research efforts in this area~\cite{Ruth:1984pz,PhysRevLett.54.693,Bane:1985uh}, most contemporary iterations of this concept are based on the so-called blowout regime~\cite{PWFA_blowout,Pukhov:2002rt}, in which the density of the driver is comparable to (or considerably higher than) that of the plasma background. In this case, the drive beam expels plasma electrons from its path in such a way that a co-moving cavity (or bubble) is created in its wake. The corresponding plasma flow is highly nonlinear, which makes the development of a rigorous, analytical theory for this regime extremely difficult. Most efforts along these lines have focused on a phenomenological treatment of the problem, using insight gained from simulations~\cite{Kostyukov_etal,Lu_pwfa,khudik_etal:2013}. Recent attempts to construct an analytical description of the blowout regime from first principles have been surprisingly successful, but the resulting expressions are only applicable in the limit of small driver charge and small driver dimensions~\cite{stupakov_BKS}. As a result, simulation codes remain the main tool for treating the PWFA problem in its general form~\cite{Nieter2004448,PhysRevSTAB.6.061301,mori_2002,HiPACE}.

Moreover, the beam loading effect associated with the presence of a realistic witness bunch and beam instability issues related to short-range wakefields (longitudinal and transverse) induced within the plasma bubble are topics of considerable importance when assessing the feasibility of the PWFA concept~\cite{Burov:2016fiw}. In this paper, we present the outline of a novel PWFA simulation code that, apart from treating the basic problem of electron acceleration, also offers a natural and convenient framework for dealing with more advanced subjects like the ones mentioned above. Starting entirely from first principles, our semi-analytical formalism is based on the assumption that we are only interested in the steady-state regime of the interaction. This quasi-static approximation implies that we neglect the longitudinal plasma non-uniformity and other fast effects associated with the injection process.
Another crucial assumption is that the PWFA configuration under consideration is axially symmetric. Restricting our analysis to the 2D case enables us to simplify our treatment while retaining most of the underlying physics. At the same time, we can readily take into account the finite size of both the driver and the witness bunches, which can have an arbitrary density profile.

Our development is organized as follows: In Section~\ref{sec:2}, we formulate the equations for the steady-state electromagnetic field excited by a beam moving with the speed of light in a uniform plasma. Section~\ref{sec:3} then derives the single particle equations of motion for the plasma electrons, while Section~\ref{sec:4} introduces some basic concepts regarding the description of the plasma flow in terms of macroparticles. The remaining details of the computational algorithm are given in Section~\ref{sec:5}, while the results of a simple numerical study of the beam loading effect are presented in Section~\ref{sec:6}. As part of the latter, we also include a comparison between our code and an existing particle-in-cell simulation tool. Our analytical derivation is completed in Section~\ref{sec:7}, where we show how our code can be adapted for the calculation of short-range wakefields inside the plasma cavity. Finally, Section~\ref{sec:8} summarizes the main results of this paper.

%
\section{Equations for the electromagnetic field}\label{sec:2}
%

In this section, we formulate the equations that describe the plasma dynamics behind a finite-size driver moving through a cold plasma with the speed of light. To start with, we assume that the driver propagates along the $z$ axis in the positive direction. As we have already mentioned, our analysis is restricted to the steady-state case, so sufficient time is assumed to have elapsed since the injection stage and the equilibrium plasma density $n_0$ is taken as uniform in space (moreover, we disregard the motion of the ions). In order to incorporate the effect of beam loading, we also assume that a finite-size, ultra-relativistic witness beam follows the driver through the plasma column at a fixed distance behind it. Both the driver and the witness bunches are assumed to be on-axis, radially symmetric and non-evolving. Following the standard convention, we normalize the time $t$ to $\omega_p^{-1}$, length to $k_p^{-1}$, and velocities to the speed of light $c$. Here, $\omega_p=ck_p$ is the plasma oscillation frequency, given by ${\omega_p} = \sqrt{4\pi{n_0}{e^2}/m} = c\sqrt {4\pi{n_0}{r_e}}$ ($r_e=e^2/mc^2$ is the classical electron radius). We also normalize momenta to $mc$, fields (electric and magnetic) to $mc\omega_p/e$, potentials (scalar and vector) to $mc^2/e$, charge densities to $n_0e$ and current densities to $en_0c$.

Assuming that all potentials and fields depend on $z$ and $t$ solely through the combination $\xi=ct-z$, which expresses the longitudinal position with respect to a fixed point of the driver, we find the following equations for the non-zero components of the (scaled) electric and magnetic fields $E_r$, $E_z$ and $B_\theta$ in the cylindrical coordinate system (for the current density, we use the relation $\vec j = -n\vec v$, where the negative electron charge is explicitly taken into account):
    \begin{subequations}\label{eq:1}
    \begin{align}\label{eq:1a}
    \frac{1}{r}
    \frac{\p}{\p r}
    rB_\theta
    &=
    \frac{\p E_z}{\p \xi}
    -
    nv_z
    -
    n_{ext}
    ,
    \\\label{eq:1b}
    \frac{\p }{\p \xi}
    (B_\theta-E_r)
    &=
    -nv_r
    ,
    \\\label{eq:1c}
    \frac{\p E_z}{\p r}
    &=
    -nv_r
    ,
    \end{align}
    \end{subequations}
where $n$ is the scaled plasma electron density, $n_{ext}=n_d+n_w$ is the total ``external'' density due to the driver and witness beams while $v_r$ and $v_z$ are the radial and longitudinal components of the electron collective velocity. We also need the equation for the divergence $\nabla\cdot\vec E$, which yields
    \begin{align}\label{eq:2}
    \frac{1}{r}
    \frac{\p}{\p r}
    rE_r
    -
    \frac{\p}{\p \xi}
    E_z
    =
    1-n
    -
    n_{ext}
    .
    \end{align}
From Eqs.~\eqref{eq:1b} and~\eqref{eq:1c}, we can introduce the pseudo-potential $\psi$ such that
    \begin{align}\label{eq:3}
    B_\theta-E_r
    =
    \frac{\p \psi}{\p r}
    ,\qquad
    E_z
    =
    \frac{\p \psi}{\p \xi}
    .
    \end{align}
The function $\psi$ is the difference between the scalar potential $\phi$ and the longitudinal component of the vector potential $A_z$, i.e. $\psi = \phi-A_z$. Combining Eq.~\eqref{eq:1a} with~\eqref{eq:2} , we obtain
    \begin{align}\label{eq:4}
    \frac{1}{r}
    \frac{\p}{\p r}
    r(B_\theta-E_r)
    &=
    n(1-v_z)
    -
    1
    ,
    \end{align}
which gives us the following equation for $\psi$:
    \begin{align}\label{eq:5}
    \frac{1}{r}
    \frac{\p}{\p r}
    r
    \frac{\p \psi}{\p r}
    &=
    n(1-v_z)-1
    .
    \end{align}
Additionally, we have an equation for the gradient $\p_\xi\psi$, namely
    \begin{align}\label{eq:6}
    \frac{\p}{\p r}
    \p_\xi\psi
    =
    -nv_r
    .
    \end{align}

We also need an equation for the azimuthal magnetic field $B_\theta$. To start with, we differentiate Eq.~\eqref{eq:1a} with respect to $r$ and substitute $\p E_z/\p r$ with the value extracted from Eq.~\eqref{eq:1c}. This yields the relation
    \begin{align}\label{eq:7}
    \frac{\p}{\p r}
    \frac{1}{r}
    \frac{\p}{\p r}
    rB_\theta
    &=
    -
    \frac{\p}{\p \xi}
    nv_r
    -
    \frac{\p}{\p r}
    nv_z
    -
    \frac{\p n_{ext}}{\p r}
    .
    \end{align}
We will discuss how Eqs.~\eqref{eq:5} and~\eqref{eq:7} are solved numerically in Section~\ref{sec:5}.

%
\section{Equations of motion for the plasma electrons}\label{sec:3}
%

In addition to the equations for the electromagnetic field derived in the previous section, we need the equations of motion for the plasma electrons. Using the relation $d\xi=(1-v_z)dt$ for the differentials $dt$ and $d\xi$, these equations can be written as
    \begin{subequations}\label{eq:8}
    \begin{align}\label{eq:8a}
    \frac{dp_r}{dt}
    =
    (1-v_z)
    \frac{d(\gamma v_r)}{d \xi}
    =
    - E_r
    +
    v_z B_\theta
    ,\\\label{eq:8b}
    \frac{dp_z}{dt}
    =
    (1-v_z)
    \frac{d(\gamma v_z)}{d \xi}
    =
    - E_z
    -
    v_r B_\theta
    ,
    \end{align}
    \end{subequations}
where $p_r=\gamma v_r$ and $p_z=\gamma v_r$ are the radial and longitudinal components of the momentum vector (note that these are now the individual particle momenta). We also have a corresponding equation for the relativistic $\gamma$ factor:
    \begin{align}\label{eq:9}
    \frac{d\gamma}{dt}
    =
    (1-v_z)
    \frac{d\gamma}{d \xi}
    =
    -E_zv_z
    -E_rv_r
    .
    \end{align}

There is an important integral of motion in this problem, given by $\gamma-p_z-\psi=\const$. This can be shown using Eqs.~\eqref{eq:8b}, \eqref{eq:9} and \eqref{eq:3}~\cite{doi:10.1063/1.872134}. Assuming that the plasma electrons are at rest at $\xi = \xi_{\rm init}$, where $\xi_{\rm init}$ is the initial (lower) $\xi$-value corresponding to the front of the driver, we obtain $\gamma - p_z - \psi = 1$ or
    \begin{align}\label{eq:10}
    1
    -
    v_z
    =
    \frac{1}{\gamma}(1+\psi)
    ,
    \end{align}
    where we have also used the fact that $\psi(r,\xi=\xi_{\rm init})=0$, a property which will be verified later on. From this, we can express $p_z$ and $\gamma$ in terms of $p_r$ and $\psi$ via the relations
    \begin{align}\label{eq:11}
    p_z
    &=
    \frac{1}{2(1+\psi)}
    [1+p_r^2-(1+\psi)^2]
    ,
    \\\nonumber
    \gamma
    &=
    \frac{1}{2(1+\psi)}
    [1+p_r^2+(1+\psi)^2]
    .
    \end{align}

Next, we re-write the equation for the radial momentum $p_r$ as
    \begin{align}\label{eq:12}
    \frac{dp_r}{d \xi}
    &=
    -
    \frac{1}{1-v_z}
    E_r
    +
    \frac{v_z }{1-v_z}
    B_\theta
    =
    -
    \frac{1}{1-v_z}
    \left(
    B_\theta-\frac{\p \psi}{\p r}
    \right)
    +
    \frac{v_z }{1-v_z}
    B_\theta
    \nonumber\\
    &=
    \frac{\gamma}{1+\psi}
    \frac{\p \psi}{\p r}
    -
    B_\theta
    ,
    \end{align}
where we have used Eqs.~\eqref{eq:3} and \eqref{eq:10}. In addition to the above, we also require an equation for the radial orbit. Starting from $dr/dt=v_r$, we obtain
    \begin{align}\label{eq:13}
    \frac{dr}{d\xi}
    =
    \frac{v_r}{1-v_z}
    =
    \frac{p_r}{1+\psi}
    .
    \end{align}

We should point out that the usefulness of $\xi=t-z$ as an independent variable instead of the time $t$ (in the particle equations) is enhanced by the fact that the former is a monotonically increasing function of the latter (recall that $d\xi/dt=1-v_z>0$). The single-particle equations of motion should be supplemented by the continuity equation for the plasma density $n$, namely
    \begin{equation}\label{eq:14}
    {\frac{1}{r}\frac{\partial }{{\partial r}}(rn{v_r})
    +
    \frac{\partial }{{\partial \xi }}n(1 - {v_z}) = 0}
    .
    \end{equation}

%
\section{Macroparticles}\label{sec:4}
%

In our computational algorithm, the plasma electrons are represented as a sum over macroparticles. Each macroparticle is characterized by the dimensionless charge $q_i$, the coordinate $r_i(\xi)$ and momenta $p_{zi}(\xi)$ and $p_{ri}(\xi)$. The coordinates and momenta satisfy the equations of motion~\eqref{eq:12} and~\eqref{eq:13}, with the fields on the right hand side of the equations taken at the positions of the particles. The plasma electron density $n$ and the corresponding currents are represented as a sum over macroparticles according to the relations
    \begin{align}\label{eq:15}
    n
    &=
    \sum_i
    \frac{q_i}{r_i(\xi)(1-v_{zi}(\xi))}
    \delta(r-r_i(\xi))
    ,
    \nonumber\\
    n(1-v_z)
    &=
    \sum_i
    \frac{q_i}{r_i(\xi)}
    \delta(r-r_i(\xi))
    ,
    \nonumber\\
    nv_r
    &=
    \sum_i
    \frac{q_i v_{ri}(\xi)}{r_i(\xi)(1-v_{zi}(\xi))}
    \delta(r-r_i(\xi))
    ,
    \end{align}
where $v_{zi}$ and $v_{ri}$ are the macroparticle velocities and the summation goes over all macroparticles in the system. The particular form of the weights in front of the delta functions in Eqs.~\eqref{eq:15} is chosen in such a way that the continuity equation~\eqref{eq:14} is automatically satisfied by the expressions given above.

The weights $q_i$ are determined by the initial coordinates of the macroparticles, $r_i(\xi _{\rm init})$. Integrating the first of Eqs.~\eqref{eq:15} from $r_1$ to $r_2$ with the weight $r$ and using  ${v_{zi}}(\xi_{\rm init}) = 0$ and $n(r,\xi_{\rm init}) = 1$ (both stemming from the fact that the plasma is at rest in front of the driver) we obtain
    \begin{equation}\label{eq:16}
    \int_{r_1}^{r_2} {d r \, r}
    =
    \frac{1}{2}(r_2^2-r_1^2)
    =
    \sum\limits_{r_1<{r_i}({\xi _{\rm init}}) < r_2} {{q_i}}
    .
    \end{equation}
To approximately satisfy this equation we choose ${q_i} = \delta {r_i}\,{r_i}({\xi_{\rm init}})$ where $\delta {r_i}$ is the separation between the particles; then the sum on the right-hand side becomes $\sum\limits_i {\delta {r_i}\,{r_i}({\xi _{\rm init}})} \approx \frac{1}{2}(r_2^2-r_1^2)$. Of course, to minimize the error stemming from this approximation one has to  use many particles with the distance between them $\delta r_i$ much smaller than the transverse size of the plasma flow.

%
\section{Calculation of the fields}\label{sec:5}
%

In this section, we focus on the solution of the equations for the plasma electromagnetic fields~\eqref{eq:5} and~\eqref{eq:7}. To start with, we decompose the magnetic field into two parts via ${B_\theta} = {{\bar B}_\theta} + B_\theta^{(0)}$, where $B_\theta^{(0)}$ is the solution of the inhomogeneous equation
    \begin{equation}\label{eq:17}
    \frac{\partial }{{\partial r}}
    \left[\frac{1}{r}\frac{{\partial (rB_\theta^{(0)})}}{{\partial r}}\right]
    =
    -
    \frac{{\partial {n_{ext}}}}{{\partial r}} \,.
    \end{equation}
This quantity, which expresses the contribution of the driver and witness beams to the total magnetic field $B_\theta$, is given in an analytical fashion by
    \begin{equation}\label{eq:18}
    {B_\theta ^{(0)}(r,\xi )
    =
    -
    \frac{1}{r}
    \int_0^r {d\hat r\,\hat r}\, {n_{ext}}(\hat r,\xi )}\,.
    \end{equation}
The equation for the part of the magnetic field due to the plasma electron flow then becomes
    \begin{align}\label{eq:19}
    \frac{\p}{\p r}
    \left[\frac{1}{r}
    \frac{\p}{\p r}
    r{\bar B}_\theta
    \right]
    &=
    -
    \frac{\p}{\p \xi}
    nv_r
    -
    \frac{\p}{\p r}
    nv_z
    .
    \end{align}
We need to solve this equation with the boundary conditions: $B_\theta=0$ at $r=0$ and $B_\theta\to 0$ when $r\to\infty$.

For the radial current $nv_r$, we use the last of Eqs.~\eqref{eq:15}, which can be written as
    \begin{align}\label{eq:20}
    nv_r
    &=
    \sum_i
    \frac{q_i p_{ri}}{r_i(1+\psi_{i})}
    \delta(r-r_i)
    ,
    \end{align}
where $\psi_i=\psi_i(\xi) = \psi(r_i(\xi),\xi)$ and we have also made use of Eq.~\eqref{eq:10}. In order to simplify the notation, we have temporarily suppressed the $\xi$-dependence of the various macroparticle quantities. For the longitudinal plasma current $nv_z$, we have the relation
    \begin{align}\label{eq:21}
    nv_z
    &=
    \sum_i
    \frac{q_iv_{zi}}{r_i(1-v_{zi})}
    \delta(r-r_i)
    =
    \sum_i
    \frac{q_i}{r_i}
    \left(
    \frac{\gamma_i}{1+\psi_{i}}
    -
    1
    \right)
    \delta(r-r_i)
    ,
    \end{align}
where we have again used Eq.~\eqref{eq:10}. The radial derivative $\p (nv_z)/\p r$ is easily calculated to be
    \begin{align}\label{eq:22}
    \frac{\p}{\p r}
    nv_z
    =
    \sum_i
    \frac{q_i}{r_i}
    \left(
    \frac{\gamma_i}{1+\psi_{i}}
    -
    1
    \right)
    \delta'(r-r_i)
    ,
    \end{align}
where the prime denotes differentiation with respect to $r$. Considerably more work is required for obtaining the longitudinal derivative $\p_\xi (nv_r)$. The end result of the corresponding calculation is
    \begin{align}\label{eq:23}
    \frac{\p}{\p \xi}
    &nv_r =
    \nonumber\\
    =& \sum_i
    \frac{q_i}{r_i}
    \left[
    \frac{\delta(r-r_i)}{1+\psi_{i}}
    \frac{dp_{ri}}{d\xi}
    -
    \frac{p_{ri}\delta'(r-r_i)}{1+\psi_{i}}
    \frac{dr_i}{d\xi}
    -
    \frac{p_{ri}\delta(r-r_i)}{(1+\psi_{i})^2}
    \left(
    \frac{dr_i}{d\xi}
    (\p_r\psi)_i
    +
    (\p_\xi \psi)_i
    \right)
    -
    \frac{p_{ri}\delta(r-r_i)}{r_i(1+\psi_{i})}
    \frac{dr_i}{d\xi}
    \right]
    \nonumber\\
    =&\sum\limits_i {\frac{{{q_i}}}{{{r_i}}}\left[ {\frac{{\delta (r - {r_i})}}{{1 + {\psi _i}}}\left( {\frac{{{\gamma _i}}}{{1 + {\psi _i}}}(\p_r\psi)_i - {\bar B_\theta } - B_\theta ^{(0)}({r_i}(\xi ),\xi )} \right) - \frac{{p_{ri}^2\delta '(r - {r_i})}}{{{{(1 + {\psi _i})}^2}}}} \right.}
    \nonumber\\
    &\left. { - \frac{{{p_{ri}}\delta (r - {r_i})}}{{{{(1 + {\psi _i})}^2}}}\left( {(\p_r\psi)_i\frac{{{p_{ri}}}}{{1 + {\psi _i}}} + ({\partial _\xi }\psi })_i \right) - \frac{{p_{ri}^2\delta (r - {r_i})}}{{{r_i}{{(1 + {\psi _i})}^2}}}} \right]\,,
    \end{align}
where use has been made of Eqs.~\eqref{eq:12} and \eqref{eq:13}. This transformation of the derivative ${\p (nv_r)}/{\p \xi}$ can be also found in Ref.~\cite{AN2013165}.

Substituting Eqs.~\eqref{eq:22} and~\eqref{eq:23} into Eq.~\eqref{eq:19} for ${\bar B}_\theta$, we observe that on the right-hand side there is a sum of terms each of which is either proportional to $\delta(r-r_i)$ or $\delta'(r-r_i)$. For $r_i<r<r_{i+1}$ (where we have tacitly assumed that the macroparticles have been labeled according to increasing radius), the plasma currents are zero and ${\bar B}_\theta$ satisfies the relation ${\partial _r}[(1/r){\partial _r}(r{\bar B_\theta })] = 0$. Thus, the magnetic field between the delta functions with labels $i$ and $i+1$ can be represented as
    \begin{align}\label{eq:24}
    \bar B_\theta
    =
    a_ir
    +
    \frac{b_i}{r}\,,
    \end{align}
where $a_i$ and $b_i$ are functions of $\xi$. Our objective now becomes to find the corresponding matching conditions at the locations of the delta functions, $r = r_i(\xi)$. These conditions are derived in Appendix~\ref{appendixA} and are formulated as a system of linear equations for $a_i$, $b_i$:
    \begin{align}\label{eq:25}
    (a_{i}-a_{i-1})r_i
    +
    \frac{b_{i}-b_{i-1}}{r_i}
    &=
    C_i
    ,
    \nonumber\\
    a_{i}-a_{i-1}
    -
    \frac{b_{i}-b_{i-1}}{r_i^2}
    &=
    A_i
    \left(
    a_{i-1}r_i
    +
    \frac{b_{i-1}}{r_i}
    \right)
    +
    B_i
    -
    \frac{C_i}{r_i}
    +
    \frac{1}{2}
    A_iC_i
    ,
    \end{align}
where the coefficients $A_i$, $B_i$ and $C_i$ are
    \begin{align}\label{eq:26}
    A_i
    &=
    \frac{q_i}{r_i}
    \frac{1}{1+\psi_{i}}
    ,\nonumber\\
    B_i
    &=
    -\frac{q_i}{r_i}
    \frac{\gamma_i}{(1+\psi_{i})^2}
    (\p_r \psi)_i
    +
    \frac{q_i}{r_i}
    \frac{p_{ri}^2}{(1+\psi_{i})^3}
    (\p_r\psi)_i
    +
    \frac{q_i}{r_i}
    \frac{p_{ri}}{(1+\psi_{i})^2}
    (\p_\xi \psi)_i
    +
    \frac{q_i p_{ri}^2}{r_i^2(1+\psi_{i})^2}
    +
    \frac{{{q_i}}}{{{r_i}}}\frac{{B_\theta ^{(0)}({r_i}(\xi ),\xi )}}{{1 + {\psi _i}}}
    ,\nonumber\\
    C_i
    &=
    \frac{q_i}{r_i}
    \frac{p_{ri}^2}{(1+\psi_{i})^2}
    -
    \frac{q_i}{r_i}
    \left(
    \frac{\gamma_i}{1+\psi_{i}}
    -
    1
    \right)
    .
    \end{align}
Eqs.~\eqref{eq:25} can also be written in matrix form,
    \begin{align}\label{eq:27}
    \begin{pmatrix}
    a_i\\b_i
    \end{pmatrix}
    =
    \begin{pmatrix}
    1+ \frac{1}{2}A_ir_i& \frac{1}{2r_i}A_i \\
    -\frac{1}{2}A_ir_i^3& 1- \frac{1}{2}A_ir_i
    \end{pmatrix}
     \begin{pmatrix}
    a_{i-1}\\b_{i-1}
    \end{pmatrix}
    +
    \begin{pmatrix}
    \frac{1}{4}
    (2B_i+A_iC_i)
    \\
    \frac{1}{4}
    r_i
    (4C_i-2B_ir_i-A_iC_ir_i)
    \end{pmatrix}.
    \end{align}
Applying the recursion rule of Eq.~\eqref{eq:27} for $1\leq i\leq N$ (where $N$ is the number of simulation particles), we can obtain the magnetic field pattern at a given $\xi$ given the pseudo-potential $\psi$ and the macroparticle positions and momenta. Apart from the condition $b_0=0$ (which avoids a singularity at $r=0$ and guarantees that $B_\theta =0 $ at $r=0$), we typically assume that $a_M=0$, which ensures a $1/r$ field decay after the last particle, and hence $B_\theta\to 0$ when $r\to\infty$.

To complete the electrodynamics-related part of our derivation, we now consider Eq.~\eqref{eq:5}. The boundary conditions for this equation are: $\psi$ is finite at $r=0$ and $\psi\to 0$ when $r\to \infty$. Substituting the second of Eqs.~\eqref{eq:15} to the right-hand side of Eq.~\eqref{eq:5}, we see that it contains terms that are proportional to $\delta(r-r_i)$. For the region between two neighboring macroparticles ($r_i<r<r_{i+1}$), we have zero plasma current so $(1/r){\partial_r}(r{\partial_r}\psi) =  - 1$. This relation yields a vacuum solution of the form $\psi  = {\bar a_i} + {\bar b_i}\ln r - {r^2}/4$, where $\bar a_i$, $\bar b_i$ are functions of $\xi$. To find relations between the adjacent coefficients $\bar a_i$, $\bar b_i$, we integrate Eq.~\eqref{eq:5} over the radius $r$ from $r_i^{-}=r_i-0$ to $r_i^{+}=r_i+0$ through the delta function $\delta(r-r_i)$. We obtain the following matching conditions for the pseudo-potential $\psi$,
    \begin{equation}\label{eq:28}
    {\partial _r}\psi (r_i^{+},\xi)
    -
    {\partial _r}\psi (r_i^{-},\xi)
    =
    \frac{{{q_i}}}{{{r_i}(\xi )}}
    ,
    \qquad
    \psi (r_i^{+},\xi) - \psi (r_i^{-},\xi) = 0\,.
    \end{equation}
With these matching conditions, we find that ${\bar a_i} + {\bar b_i}\ln {r_i} = {\bar a_{i - 1}} + {\bar b_{i - 1}}\ln {r_i}$ and ${\bar b_i} - {\bar b_{i - 1}} = {q_i}$. For $0 < r < {r_1}$ (i.e. for radii smaller than that of the first macroparticle), we have $\psi  = {\bar a_0} - {r^2}/4$ since $\bar b_0$ must be equal to zero in order to avoid a singularity at $r=0$. Using the above, we can add potential terms over successive delta functions in order to construct the solution for $\psi$. The end result is $\psi(r,\xi)=\psi(0,\xi)+\Delta\psi(r,\xi)$, where
    \begin{equation}\label{eq:29}
    \Delta\psi (r,\xi)
    =
    \,\sum\nolimits_{{r_i(\xi)} < \,\,r}
    {\,{q_i}\ln \left[\frac{r}{r_i(\xi)}\right]}  - \frac{{{r^2}}}{4}
    \end{equation}
and
    \begin{equation}\label{eq:30}
    \psi (0,\xi)
    =
    {\bar a_0}
    =
    -
    \Delta\psi(r_N(\xi),\xi)
    =
    -
    \sum\nolimits_{i} {\,{q_i}
    \ln \left[\frac{r_N(\xi)}{r_i(\xi)}\right]}  + \frac{{r_N^2(\xi)}}{4}\,.
    \end{equation}
Here, we have adopted the condition that the pseudo-potential $\psi$ is always zero at the location of the last macroparticle, i.e. $\psi(r_N(\xi),\xi) = 0$. For large enough radial coordinate of the last electron in the system, $r_N(\xi)$, this condition approximates the boundary condition $\psi\to 0$ when $r\to\infty$. From Eqs.~\eqref{eq:29} and \eqref{eq:30}, one can easily derive the partial derivatives of $\psi$ that are needed for evaluating the $A_i$, $B_i$, $C_i$ coefficients discussed earlier. In particular, we have
    \begin{equation}\label{eq:31}
    {\partial _r}\psi (r,\xi) = {\partial _r}\Delta\psi (r,\xi) =\frac{1}{r}\sum\nolimits_{{r_i(\xi)} < \,\,r} {{q_i}\,}  - \frac{r}{2}
    \end{equation}
for the radial derivative and ${\partial_\xi}\psi(r,\xi)={\partial_\xi}\Delta\psi(r,\xi)-{\partial_\xi}\Delta\psi(r_N(\xi),\xi)$ for its longitudinal counterpart, where
    \begin{equation}\label{eq:32}
    {\partial_\xi}\Delta\psi(r,\xi) =  - \sum\nolimits_{{r_i(\xi)} < \,\,r} {\,\frac{{{q_i}}}{{{r_i}}}} \frac{{{p_{ri}}}}{{1 + {\psi _i}}}
    ,
    \end{equation}
and we have assumed that $d{r_N}(\xi)/d\xi  = 0$. Incidentally, it is worth mentioning that one of the $\xi$-derivatives mentioned above is directly related to the calculation of the on-axis longitudinal electric field, one of the main figures of merit for the PWFA. Specifically, the force per unit charge for an on-axis electron is given by ${F_z}/e=-{E_z}(0,\xi)={\partial_\xi}\psi(0,\xi)=-{\partial_\xi}\Delta\psi({r_N},\xi)$, a result which is particularly useful for numerical calculations. Lastly, we point out that combining Eqs.~\eqref{eq:16} and \eqref{eq:31} yields ${\partial _r}\psi (r,\xi_{\rm init}) = 0$, which means that ${\psi (r,\xi_{\rm init}) = \psi ({r_N},\xi_{\rm init}) = 0}$. This verifies our earlier claim that the pseudo-potential is zero for all $r$ at the front of the driver.

Eqs.~\eqref{eq:18}, \eqref{eq:24}, \eqref{eq:27} and \eqref{eq:26}, along with Eqs.~\eqref{eq:29}-\eqref{eq:32}, form the basic results of our field analysis. Their main feature is that a knowledge of the macroparticle positions and momenta at a given $\xi$ suffices for a complete description of the electromagnetic fields at that particular (relative) longitudinal position. This is also reflected in the absence of any derivatives with respect to $\xi$ on the LHS of Eqs.~\eqref{eq:19} and \eqref{eq:5}, a property which prevents the coupling of different $\xi$-values. As a result, one can readily make use of the single-particle equations of motion (i.e. Eqs.~\eqref{eq:11}-\eqref{eq:13}) in order to propagate the particle positions and momenta to $\xi+\delta\xi$. To avoid any confusion, we rewrite the latter in our updated (macroparticle) notation as
\begin{equation}\label{eq:33}
\frac{{d{p_{ri}}}}{{d\xi }} = \frac{{{\gamma _i}}}{{1 + {\psi _i}}}{({\partial _r}\psi )_i} - {\bar B_\theta}({r_i},\xi) - {B_\theta^{(0)}}({r_i},\xi)\,\,\,,\,\,\,\frac{{d{r_i}}}{{d\xi }} = \frac{{{p_{ri}}}}{{1 + {\psi _i}}}
\end{equation}
and
\begin{align}\label{eq:34}
    p_{zi}
    &=
    \frac{1}{2(1+\psi_i)}
    [1+p_{ri}^2-(1+\psi_i)^2]
    ,
    \\\nonumber
    \gamma_i
    &=
    \frac{1}{2(1+\psi_i)}
    [1+p_{ri}^2+(1+\psi_i)^2]
    .
\end{align}
In our algorithm, we use a fourth-order Runge-Kutta technique to numerically integrate the equations of motion for the plasma electrons. Another important point is that, when calculating discontinuous quantities like $\bar B_\theta$, ${\partial_r}\psi$ or ${\partial_\xi}\psi$ at the location of a macroparticle (i.e. at the actual location of the discontinuity), we use the arithmetic mean of the two limiting values. This is simply due to the fact that the simulation macroparticles really represent charged sheets formed by the nonlinear plasma flow. In any event, knowing the new positions and momenta enables us to calculate the fields at the updated location. Thus, a full solution of the steady-state PWFA problem (including details such as the shape of the cavity) can be constructed through this systematic, step-by-step process.

We have implemented our algorithm in the computer code PLEBS (PLasma-Electron Beam Simulations) using the MATLAB programming environment. In the next section, we will compare the results obtained with this code against QuickPic simulations.

%
\section{Beam loading study}\label{sec:6}

So far, we have kept the density profiles of the driver and witness beams entirely general. Here, we choose to specialize to the case where both beams have a Gaussian profile in the transverse and longitudinal directions. In doing so, we find it most convenient to start from the original quantities and then switch to their dimensionless counterparts. To begin with, we assume that the density of the drive beam is given by
\begin{equation}\label{eq:35}
{n_d}(r,\xi) = {n_{d0}}\exp \left( { - \frac{{{r^2}}}{{2\sigma_{rd}^2}}} \right)\exp \left( { - \frac{{{(\xi-\xi_d)^2}}}{{2\sigma_{\xi d}^2}}} \right)\,,
\end{equation}
where $n_{d0}$ is the peak volume density, $\sigma_{rd}$ and $\sigma_{\xi d}$ are (respectively) the transverse and longitudinal rms beam sizes and $\xi_d$ is the location of the drive beam centroid. Regarding the origin of $\xi$, we select $\xi_{\rm init}=0$ and require this initial value to correspond to the front (or head) of the drive beam. This, in turn, is virtually guaranteed if we choose $\xi_d=N_0\sigma_{\xi d}$, where $N_0\geq3$. The current profile of the drive beam is
\begin{equation}\label{eq:36}
{I_d}(\xi ) = ce\int {{n_d}(r,\xi )2\pi rdr = {I_{d0}}} \exp \left( { - \frac{{{{(\xi  - {\xi _d})}^2}}}{{2\sigma _{\xi d}^2}}} \right)\,,
\end{equation}
where ${I_{d0}}=2\pi ce{n_{d0}}\sigma_{rd}^2$ is the peak current. The total charge contained in the drive beam is ${Q_d} = e{N_d} = \sqrt {2\pi} {I_{d0}}({\sigma _{\xi d}}/c)$, where ${N_d} = {n_{d0}}{(2\pi )^{3/2}}\sigma _{rd}^2{\sigma _{\xi d}}$ is the total number of electrons. As far as the witness beam is concerned, we have an analogous Gaussian expression, namely
\begin{equation}\label{eq:37}
{n_w}(r,\xi) = {n_{w0}}\exp \left( { - \frac{{{r^2}}}{{2\sigma_{rw}^2}}} \right)\exp \left( { - \frac{{{(\xi-\xi_w)^2}}}{{2\sigma_{\xi w}^2}}} \right)\,.
\end{equation}
The corresponding current profile is given by
\begin{equation}\label{eq:38}
{I_w}(\xi ) = {I_{w0}} \exp \left( { - \frac{{{{(\xi  - {\xi _w})}^2}}}{{2\sigma _{\xi w}^2}}} \right)\,,
\end{equation}
with ${I_{w0}}=2\pi ce{n_{w0}}\sigma_{rw}^2$, ${Q_w} = e{N_w} = \sqrt {2\pi} {I_{w0}}({\sigma _{\xi w}}/c)$ etc. At this point, we switch back to the dimensionless variables employed in most of this text (we recall that volume densities are scaled with respect to the plasma density $n_0$, while lengths are normalized with respect to the plasma skin depth $k_p^{-1}$). The scaled drive beam density is
\begin{equation}\label{eq:39}
{{\tilde n}_d} = {{\tilde n}_{d0}}\exp \left( { - \frac{{{{\tilde r}^2}}}{{2\tilde \sigma_{rd}^2}}} \right)\exp \left( { - \frac{{{{(\tilde\xi-\tilde\xi_d) }^2}}}{{2\tilde\sigma_{\xi d}^2}}} \right)\,,
\end{equation}
where the tildes (temporarily) denote the scaled notation and ${{\tilde n}_{d0}}$ is given by ${{\tilde n}_{d0}}=\sqrt{2/\pi}\,\nu_d/(\tilde\sigma_{rd}^2{{\tilde\sigma}_{\xi d}})$. Here, $\nu_d\equiv{N_d}{r_e}{k_p}$ is the dimensionless charge of the driver. This important quantity can be shown to be proportional to the ratio of the number of particles in the beam to the number of plasma particles within a sphere of radius $k_p^{-1}$~\cite{stupakov_BKS}. Needless to say, an entirely analogous expression can be obtained for the witness beam.

Dropping the tildes from now on, we use the above results and Eq.~\eqref{eq:18} in order to determine the external field $B_\theta^{(0)}$ for our case. The resulting analytical expression is
\begin{align}\label{eq:40}
B_\theta ^{(0)}(r,\xi ) =  &- {n_{d0}}\exp \left( { - \frac{{{(\xi-\xi_d)^2}}}{{2\sigma _{\xi d}^2}}} \right)\frac{{\sigma _{rd}^2}}{r}[1 - \exp \left( { - \frac{{{r^2}}}{{2\sigma _{rd}^2}}} \right)]
\\\nonumber
&- {n_{w0}}\exp \left( { - \frac{{{(\xi-\xi_w)^2}}}{{2\sigma _{\xi w}^2}}} \right)\frac{{\sigma _{rw}^2}}{r}[1 - \exp \left( { - \frac{{{r^2}}}{{2\sigma _{rw}^2}}} \right)]
.
\end{align}

With the above relation, we have everything that we need in order to study the beam loading effect for a Gaussian drive/witness beam configuration. For the drive beam, the parameters we have used are $\sigma_{rd}=2$ $\mu$m, $\sigma_{\xi d}=20$ $\mu$m, $I_{d0}=2$ kA (so that $Q_d=334$ pC) and $\xi_d=3\sigma_{rd}$. As far as the witness beam is concerned, we assume that $\sigma_{rw}=2$ $\mu$m, $\sigma_{\xi w}=6$ $\mu$m, $\xi_w=130$ $\mu$m and $I_{w0}=2.125/2.55$ kA (we have examined two different cases, corresponding to a charge of 106/127 pC). The plasma density $n_0$ is taken to be $7\times10^{16}\,\rm cm^{-3}$, which leads to a skin depth of $k_p^{-1}=20$ $\mu$m. For these parameters, the dimensionless charge of the drive beam is $\nu_d=0.293$, while that of the witness bunch is $\nu_w=0.09/0.11$. Moreover, we have $n_{d0}\approx23.7$ and $n_{w0}\approx25.2/30.2$. The fact that these scaled density values are much larger than unity is a typical characteristic of the blowout regime of the PWFA.

\begin{figure}[ht!]
   \centering
   \includegraphics[width=197pt]{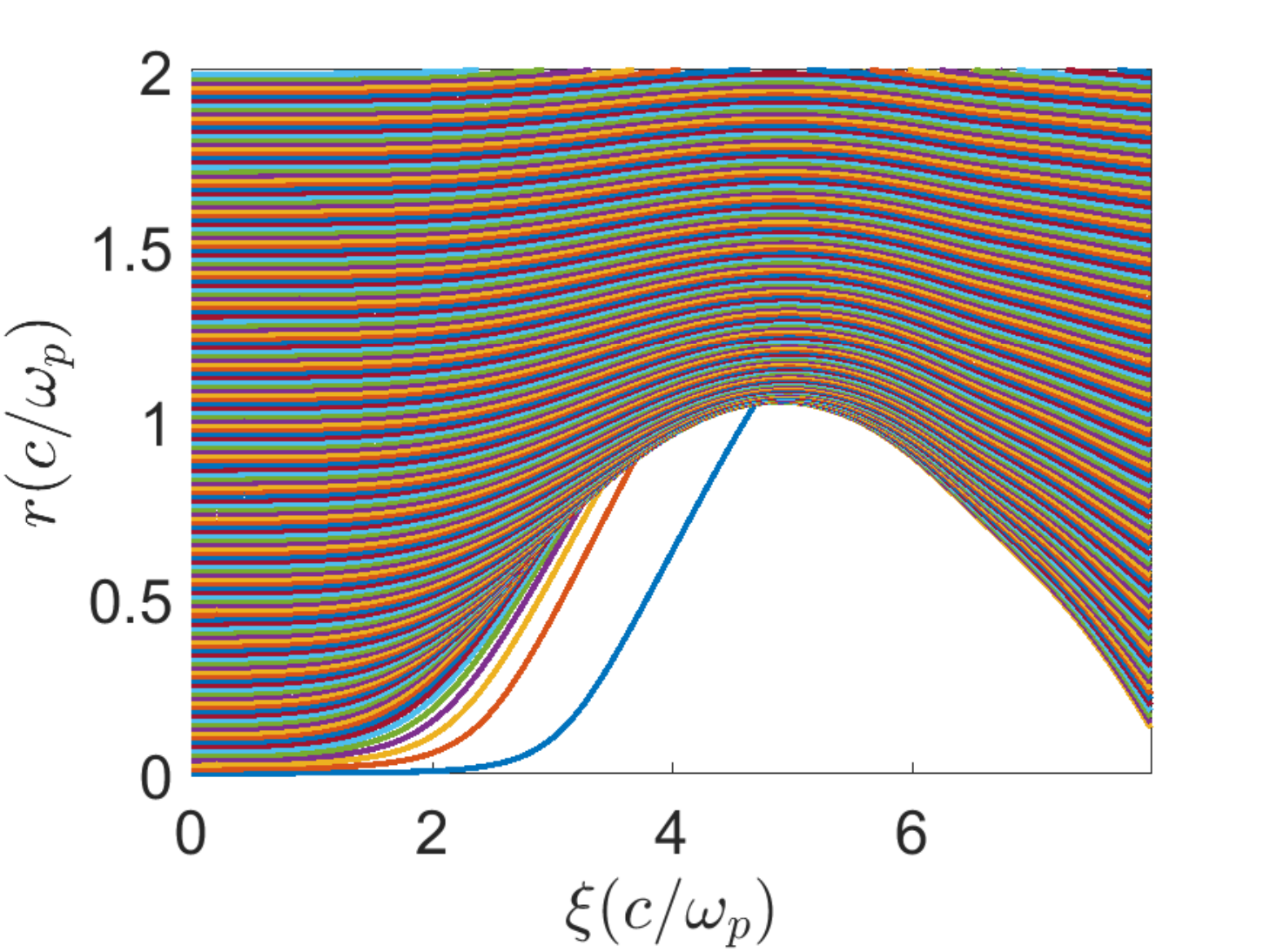}
   \includegraphics[width=197pt]{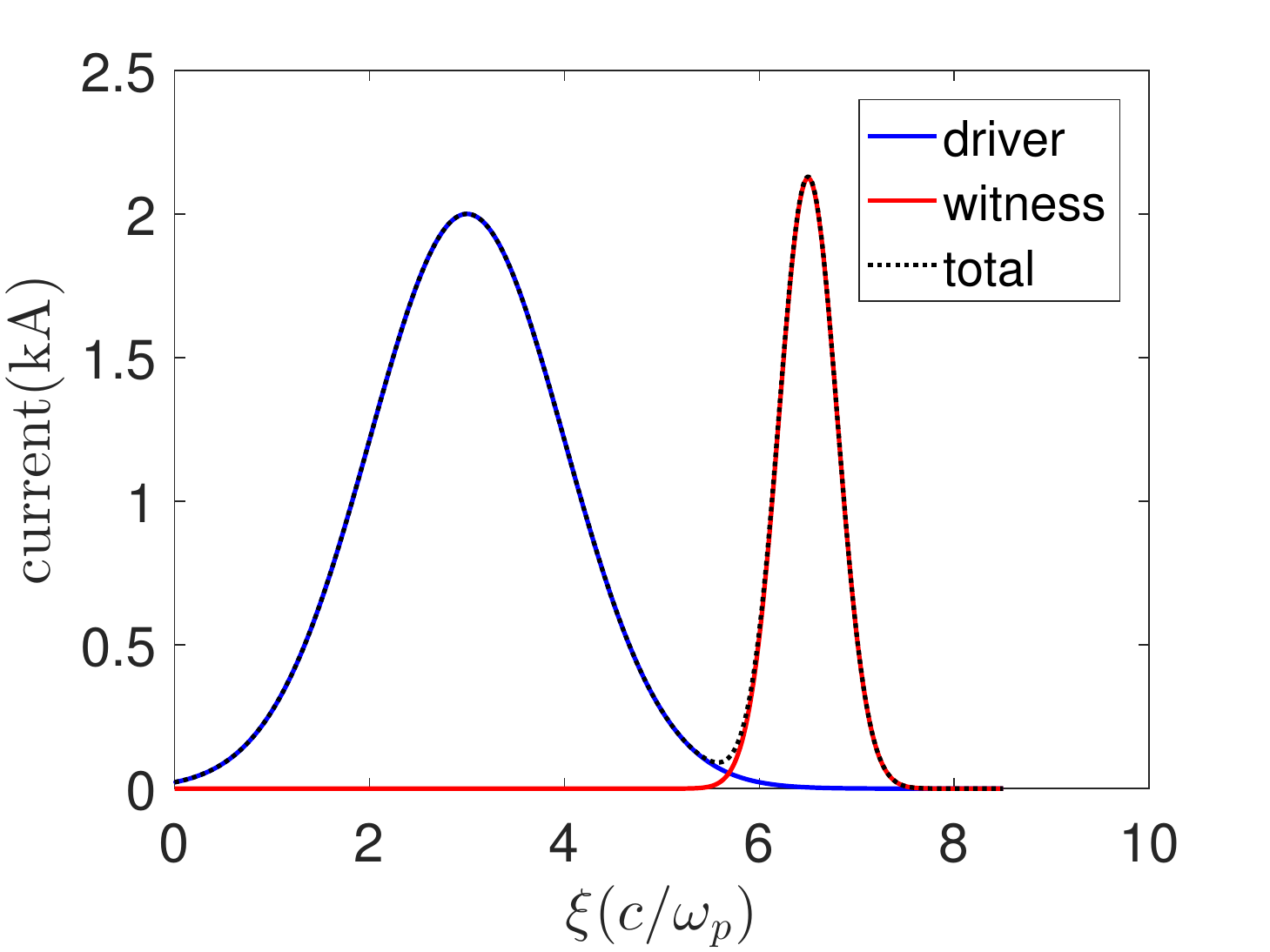}
   \caption{Right figure: current profiles of the Gaussian drive and witness beams. Left figure: electron trajectories in the $r-\xi$ plane. The formation of a plasma cavity (or bubble) is evident.}
   \label{fig:1}
\end{figure}

\begin{figure}[hb!]
   \centering
   \includegraphics[width=197pt]{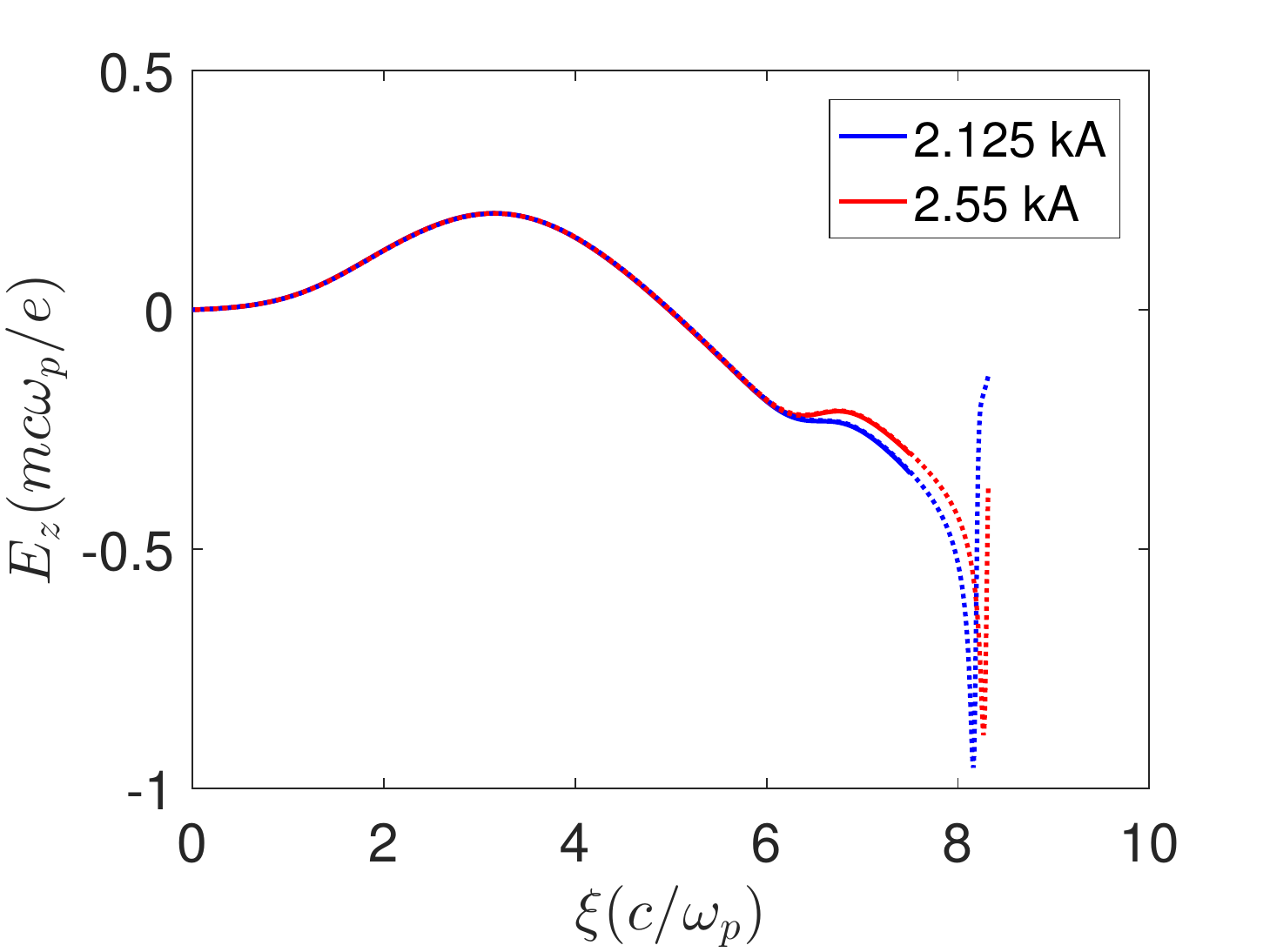}
   \includegraphics[width=197pt]{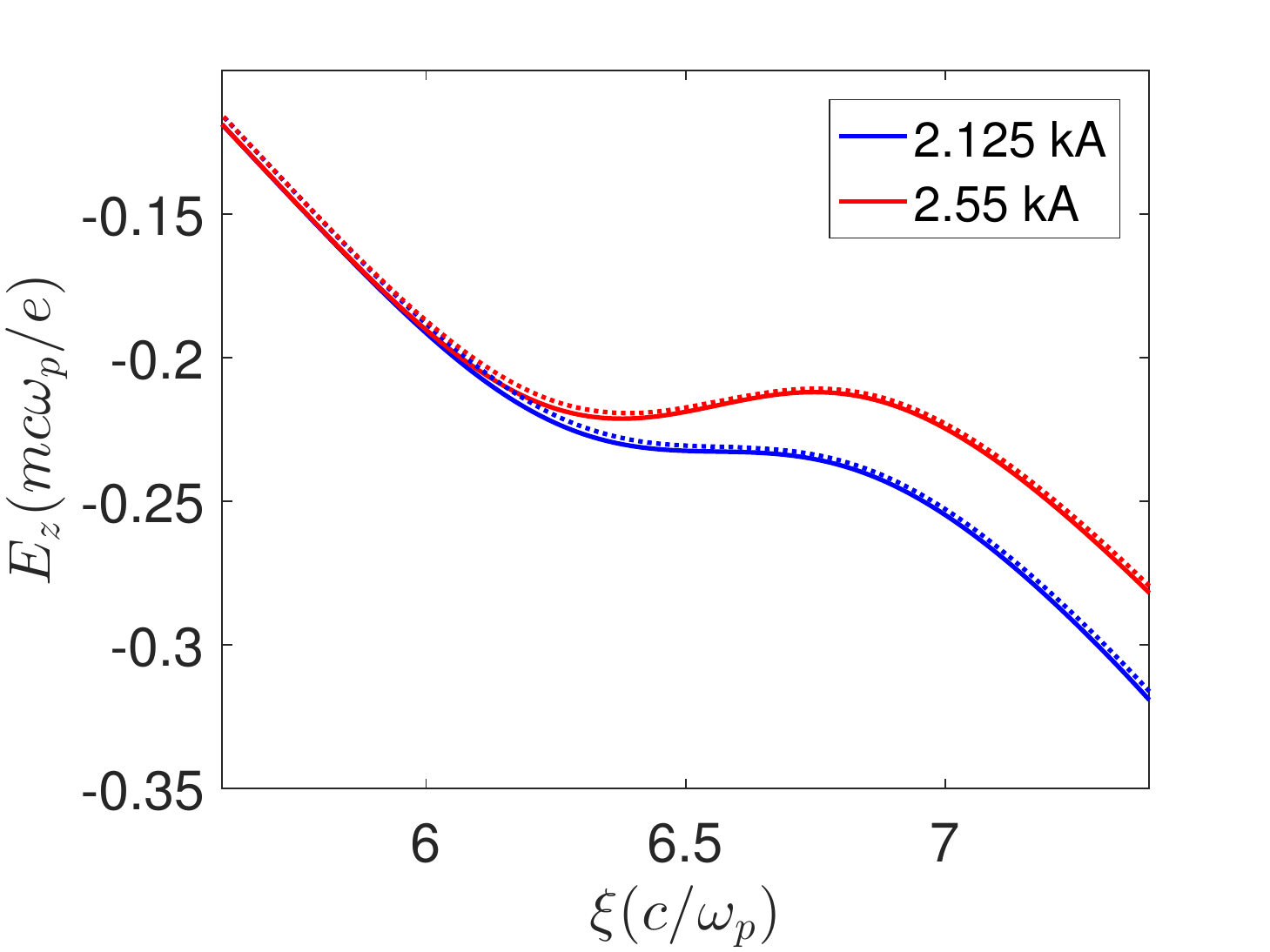}
   \caption{On-axis longitudinal electric field $E_z(0,\xi)$ as a function of $\xi$ (scaled units). We show the results form our code (solid lines) versus QuickPic data (dashed lines) for two different values of the witness beam current. The right figure focuses on the immediate neighborhood of the witness beam, while the left figure shows the full range of $\xi$.}
   \label{fig:2}
\end{figure}

The current profiles for the drive and witness bunches are illustrated in the right graph of Fig.~\ref{fig:1} (for a witness current of $2.125$ kA). In the left graph of the same figure, we plot a large number of electron trajectories whose initial radii $r_i(\xi=0)$ are uniformly spaced on the $r$-axis. More specifically, our simulation run makes use of 2000 macroparticles with $0<r_i(\xi=0)<5$, though we only plot 400 representative trajectories. The formation of a bubble is evident, with a maximum radius $r_{\rm cav}\approx1$ (about 20 $\mu$m) and a length $\xi_{\rm cav}\approx8$ (or 160 $\mu$m). In Fig.~\ref{fig:2}, we plot the on-axis, longitudinal electric field $E_z(0,\xi)$ as a function of $\xi$ for the two cases of witness beam current. The solid lines correspond to results from PLEBS, and the dashed lines represent data from QuickPic~\cite{HUANG2006658}, a popular particle-in-cell PWFA simulation tool. The left graph shows the comparison for the full range of $\xi$ (up to the bubble collapse), while the right graph zooms in on the location of the witness bunch. According to a well-known pattern, the electric field is initially decelerating but changes sign as one moves further away from the driver. Moreover, the beam loading effect only influences the field in the neighborhood of the witness beam. Overall, very good agreement is observed between the two approaches, while our technique appears to offer an advantage in terms of computation speed. In particular, our MATLAB-based code only takes a few minutes to run on a simple desktop environment, whereas QuickPic requires a parallel setup with 128 cores in order to achieve the same performance.

\section{Wakefield calculation}\label{sec:7}
%

In this section, our aim is to demonstrate that the algorithm which was described in the previous parts of this paper can also be used to calculate the short-range wakefields induced within the plasma bubble. For the basic definitions, notation and analytical results regarding the topic of wakefields in a PWFA context, we rely on work presented elsewhere~\cite{GSWP}. To start with, we turn to the problem of the wakefields excited by an on-axis point charge $q$ located inside an axisymmetric plasma cavity at the longitudinal position $\xi=\xi_0$. Our goal is to calculate the longitudinal wakefield immediately behind the charge, at $\xi=\xi_0^+$. We denote by $\tilde E_r(r,\xi)$, $\tilde E_z(r,\xi)$ and $\tilde B_\theta(r,\xi)$ the components of the modified field in the presence of the point charge $q$. As is shown in~\cite{GSWP}, the transverse field components can be expressed as
\begin{subequations}\label{eq:41}
    \begin{align}\label{eq:41a}
    \tilde E_r
    =
    E_r(r,\xi)
    +
    \Delta E_r(r,\xi) h(\xi-\xi_0)
    -
    D(r,\xi_0)\delta(\xi-\xi_0)
    ,
    \\
    \tilde B_\theta\label{eq:41b}
    =
    B_\theta(r,\xi)
    +
    \Delta B_\theta(r,\xi) h(\xi-\xi_0)
    -
    D(r,\xi_0)\delta(\xi-\xi_0)
    ,
    \end{align}
    \end{subequations}
where $E_r$, $E_z$ and $B_\theta$ are the original components, $h(\xi)$ is the step function equal to one for positive arguments and zero otherwise, $\Delta E_r$ and $\Delta B_\theta$ denote the change of the field due to the charge $q$, and the terms with the delta function represent a shock wave characterized by the radial profile $D(r,\xi_0)$. The latter quantity satisfies the relation
\begin{align}\label{eq:42}
    \frac{\p}{\p r}
    \left[\frac{1}{r}
    \frac{\p}{\p r}
    rD(r,\xi_0)\right]
    &=
    \frac{ n(r,\xi_0)}{ \gamma(r,\xi_0)}
    D(r,\xi_0)
    .
    \end{align}
The longitudinal electric field $E_z$ also exhibits a discontinuity, which is expressed by
    \begin{align}\label{eq:43}
    \tilde E_z(r,\xi)
    =
    E_z(r,\xi)
    +
    \Delta E_z(r,\xi)
    h(\xi-\xi_0)
    .
    \end{align}
The jump $\Delta E_z(r,\xi)$ represents the longitudinal wake generated by the charge $q$ immediately behind it. It can be related to the $D$-function by means of the relation
    \begin{align}\label{eq:44}
    \Delta E_z(r,\xi_0)
    =
    -
    \frac{1}{r}
    \frac{\p}{\p r}
    rD(r,\xi_0)
    .
    \end{align}
Thus, calculating the longitudinal wakefield involves solving Eq.~\eqref{eq:42} for a given (that is, known from a previous calculation) plasma density $n(r,\xi)$. Using the first of Eqs.~\eqref{eq:15}, we can re-write~\eqref{eq:42} as
    \begin{align}\label{eq:45}
    \frac{\p}{\p r}
    \frac{1}{r}
    \frac{\p}{\p r}
    rD
    &=
    D
    \sum_i
    \frac{q_i}{r_i\gamma_i(1-v_{zi})}
    \delta(r-r_i)
    =
    D
    \sum_i
    A_i
    \delta(r-r_i)
   ,
    \end{align}
where $A_i$ has been defined in Eq.~\eqref{eq:26} and we have used Eq.~\eqref{eq:10}. The above equation can be solved using an iterative technique analogous to the one we employed in solving Eq.~\eqref{eq:19} (we note the similarity between these two equations, which have essentially identical left-hand sides). Specifically, since the RHS of~\eqref{eq:45} is zero in the intervals $r_i<r<r_{i+1}$ (for $i=1,\ldots,N-1$), we seek solutions of the form $D = \hat a_i r+\hat b_i/r$, where $\hat a_i$, $\hat b_i$ only depend on $\xi$. Furthermore, we have $D = \hat a_0 r+2\nu_q/r$ for $r<r_1$ and $D = \hat b_N/r$ for $r>r_N$. Here, $\nu_q= qr_ek_p/e$ is the scaled charge and we have also taken into account the asymptotic behavior of the shock-like field induced by a point charge in a uniform plasma ($E_r\propto\nu_q/r$ for $r\rightarrow0$, according to~\cite{barov_2004}). Lastly, our choice of $\hat a_N=0$ ensures that $D\rightarrow0$ at $r\rightarrow\infty$. At the location of each delta function, we have the matching conditions of continuity and derivative jump for $D$, given by
\begin{equation}\label{eq:46}
D(r_i^+,\xi) = D(r_i^-,\xi) \,\,\,\,,\,\,\, {\partial_r}D(r_i^+,\xi) - {\partial_r}D(r_i^-,\xi) = {A_i}D(r_i^-,\xi)\,.
\end{equation}
These conditions yield the relations
\begin{align}\label{eq:47}
    \hat a_i r_i+\frac{\hat b_i}{r_i}
    =
    \hat a_{i-1} r_i+\frac{\hat b_{i-1}}{r_i}
    ,\qquad
    \hat a_i -\frac{\hat b_i}{r_i^2}
    =
    \hat a_{i-1} -\frac{\hat b_{i-1}}{r_i^2}
    +
    A_i
    \left(
    \hat a_{i-1} r_i+\frac{\hat b_{i-1}}{r_i}
    \right)
    ,\qquad
    i=1,\ldots,N
    \end{align}
with $\hat b_0=2\nu_q$ and $\hat a_N=0$. In matrix notation, the solution of this system is
    \begin{align}\label{eq:48}
    \begin{pmatrix}
    \hat a_i   \\
    \hat b_i
	\end{pmatrix}
	=
	\hat M_i
    \begin{pmatrix}
    \hat a_{i-1}   \\
    \hat b_{i-1}
	\end{pmatrix}\,,
    \end{align}
where the transfer matrix $\hat M_i$ is given by
    \begin{align}\label{eq:49}
	\hat M_i
	=
    \begin{pmatrix}
    1+\frac{1}{2} A_ir_i & \frac{1}{2r_i} A_i \\
    -\frac{1}{2} A_ir_i^3& 1-\frac{1}{2} A_ir_i
	\end{pmatrix}\,.
    \end{align}
We note that this particular matrix is also present in Eq.~\eqref{eq:27}. Combining these manipulations, we obtain
    \begin{align}\label{eq:50}
    \begin{pmatrix}
    \hat a_N   \\
    \hat b_N
	\end{pmatrix}
	=
	\hat M
    \begin{pmatrix}
    \hat a_0   \\
    \hat b_0
	\end{pmatrix}
    =
    \begin{pmatrix}
    \hat M_{11}&\hat M_{12}   \\
    \hat M_{21}&\hat M_{22}
	\end{pmatrix}
    \begin{pmatrix}
    \hat a_0   \\
    \hat b_0
	\end{pmatrix}
    \,,
    \end{align}
where $\hat M=\hat M_N\hat M_{N-1}\ldots \hat M_1$ is a cumulative matrix. Recalling that $\hat b_0=2\nu_q$ and $\hat a_N=0$, we obtain ${\hat a_0} =  - 2\nu_q{\hat M_{12}}/{\hat M_{11}}$. Combining Eq.~\eqref{eq:44} with the small-radius expression for $D$ ($D=\hat a_0r+2\nu_q/r$), we find that the on-axis value of the longitudinal wakefield is given by
    \begin{align}\label{eq:51}
    \Delta E_z (0,\xi)
    =
    -2\hat a_0
    =
    \frac{4\nu_q \hat M_{12}}{\hat M_{11}}\,.
    \end{align}

\begin{figure}[ht!]
   \centering
   \includegraphics[width=197pt]{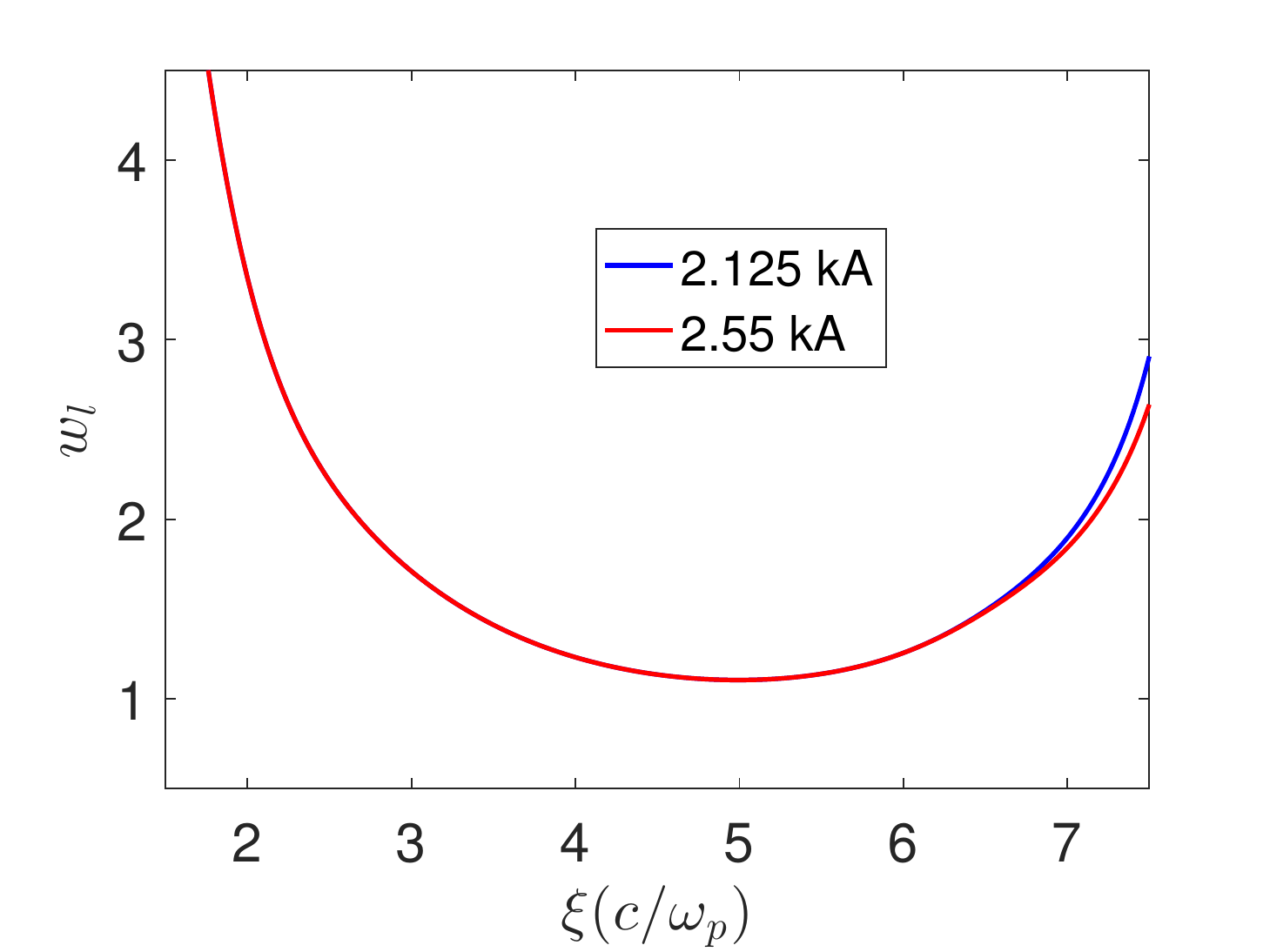}
   \includegraphics[width=197pt]{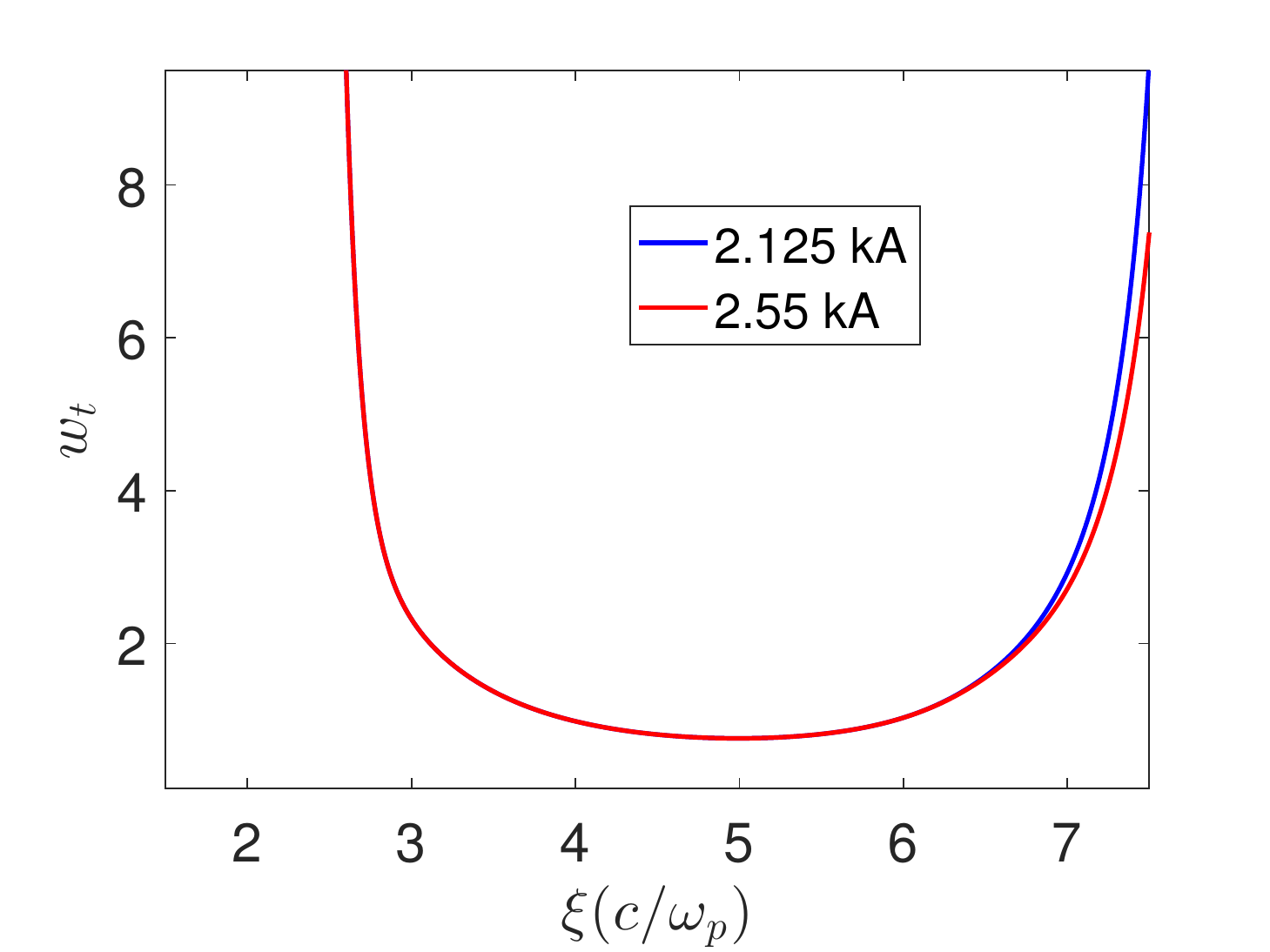}
   \caption{Scaled longitudinal and transverse wakefields $w_l=-2\hat a_0/\nu_q$ and $w_t=-4\tilde a_0/d$ as functions of $\xi$. The results shown here correspond to the parameter set used in the beam loading study of Section~\ref{sec:6}.}
   \label{fig:3}
\end{figure}

For the transverse wakefield calculation, the configuration of the leading point charge is somewhat different. In particular, the point charge $q$ is now off-axis, moving with a transverse offset $\vec{a}$ inside the plasma cavity (we assume $\left|{\vec{a}}\right|\ll k_p^{-1}$). This time, the perturbed electromagnetic field, which now lacks axial symmetry, is given by
    \begin{align}\label{eq:52}
    \tilde{\vec E}_\perp
    &=
    \vec E_\perp
    +
    \Delta \vec E_\perp(x,y,\xi)
    h(\xi-\xi_0)
    -
    \vec D(x,y,\xi)\delta(\xi-\xi_0)
    \nonumber\\
    \tilde{\vec B}_\perp
    &=
    \vec B_\perp
    +
    \Delta \vec B_\perp(x,y,\xi)
    h(\xi-\xi_0)
    -
    \hat{\vec z}\times\vec D(x,y,\xi)\delta(\xi-\xi_0)
    \nonumber\\
    \hat E_z
    &=
    E_z
    +
    \Delta E_zh(\xi-\xi_0)
    .
    \end{align}
Assuming that $\vec{a}=a\hat x$, the components of the shock profile $\vec{D}$ (in cylindrical coordinates) are
    \begin{align}\label{eq:53}
    \begin{pmatrix}
	D_r  \\
	D_\theta
	\end{pmatrix}
    =
    u(r,\xi)
    \begin{pmatrix}
	\cos\theta  \\
	\sin\theta
	\end{pmatrix}
    \,,
    \end{align}
where the radial profile $u(r,\xi)$ satisfies the relation
\begin{align}\label{eq:54}
    \frac{{{\partial ^2}u}}{{\partial {r^2}}} + \frac{1}{r}\frac{{\partial u}}{{\partial r}} - \frac{{4u}}{{{r^2}}}
    =
    u
    \frac{n(r,\xi)}{\gamma(r,\xi)}\,.
    \end{align}
Our objective is to solve this equation using our matrix technique. Substituting the expression for $n$ from Eqs.~\eqref{eq:15}, we can re-write~\eqref{eq:54} as
 \begin{align}\label{eq:55}
    \frac{{{\partial ^2}u}}{{\partial {r^2}}} + \frac{1}{r}\frac{{\partial u}}{{\partial r}} - \frac{{4u}}{{{r^2}}}
    =
    u
    \sum_i
    A_i
    \delta(r-r_i)
    \,.
    \end{align}
In the intervals $r_i<r<r_{i+1}$, the right-hand side is zero and we can easily show that the appropriate vacuum solution for $u$ is $u = \tilde a_i r^2+\tilde b_i/r^2$, $\tilde a_i$ and $\tilde b_i$ being functions of $\xi$ ($i=1,\ldots,N-1$). For $r<r_1$, the correct behavior is $u = \tilde a_0 r^2+2d/r^2$ (where $d=qa$ is the dipole moment, according to~\cite{GSWP}), while $u = \tilde b_N/r^2$ for $r>r_N$ ($\tilde a_N=0$, so that $u\rightarrow0$ when $r\rightarrow\infty$). The matching conditions for $u$ at $r=r_i$ are actually the same as those of~\eqref{eq:46}, namely
\begin{equation}\label{eq:56}
u(r_i^+,\xi) = u(r_i^-,\xi) \,\,\,\,,\,\,\, {\partial_r}u(r_i^+,\xi) - {\partial_r}u(r_i^-,\xi) = {A_i}u(r_i^-,\xi)\,.
\end{equation}
Given the analytical expression for $u$, these conditions yield
    \begin{align}\label{eq:57}
    \tilde a_i r_i^2+\frac{\tilde b_i}{r_i^2}
    =
    \tilde a_{i-1} r_i^2+\frac{\tilde b_{i-1}}{r_i^2}
    ,\qquad
    2\tilde a_i r_i -\frac{2\tilde b_i}{r_i^3}
    =
    2\tilde a_{i-1} r_i -\frac{2\tilde b_{i-1}}{r_i^3}
    +
    A_i
    \left(
    \tilde a_{i-1} r_i^2+\frac{\tilde b_{i-1}}{r_i^2}
    \right)
    ,\qquad
    i=1,\ldots,N
    \end{align}
with $\tilde b_0=2d$ and $\tilde a_N=0$. The above results can be cast into a recurrence relation:
    \begin{align}\label{eq:58}
    \begin{pmatrix}
    \tilde a_i   \\
    \tilde b_i
	\end{pmatrix}
	=
	\tilde M_i
    \begin{pmatrix}
    \tilde a_{i-1}   \\
    \tilde b_{i-1}
	\end{pmatrix}\,,
    \end{align}
where the transfer matrix $\tilde M_i$ is
    \begin{align}\label{eq:59}
	\tilde M_i
	=
    \begin{pmatrix}
    1+\frac{1}{4} A_ir_i & \frac{1}{4r_i^3} A_i \\
    -\frac{1}{4} A_ir_i^5& 1-\frac{1}{4} A_ir_i
	\end{pmatrix}\,.
    \end{align}
Combining these matrix manipulations, we obtain
    \begin{align}\label{eq:60}
    \begin{pmatrix}
    \tilde a_N   \\
    \tilde b_N
	\end{pmatrix}
	=
	\tilde M
    \begin{pmatrix}
    \tilde a_0   \\
    \tilde b_0
	\end{pmatrix}
    =
    \begin{pmatrix}
    \tilde M_{11}&\tilde M_{12}   \\
    \tilde M_{21}&\tilde M_{22}
	\end{pmatrix}
    \begin{pmatrix}
    \tilde a_0   \\
    \tilde b_0
	\end{pmatrix}
    \,,
    \end{align}
where $\tilde M=\tilde M_N\tilde M_{N-1}\ldots \tilde M_1$ is the new cumulative matrix. Recalling that $\tilde b_0=2d=2qa$ and $\tilde a_N=0$, we obtain ${\tilde a_0}=-2d{\tilde M_{12}}/{\tilde M_{11}}$. The coefficient $\tilde a_0$ has the following physical meaning: the transverse force per unit charge $F_x$ (scaled by $mc\omega_p/e$) acting on a trailing point charge at $\xi>\xi_0$ can be shown to be $F_x=-4\tilde a_0(\xi-\xi_0)$. Thus, $\tilde a_0$ can be thought of as a measure of the transverse wake. A basic numerical illustration of the variation of the wakefields along the bubble is given in Fig.~\ref{fig:3}. Specifically, we plot the quantities $w_l=\Delta E_z(0,\xi)/\nu_q=4\hat M_{12}/\hat M_{11}$ and $w_t=-4\tilde a_0/d=8\tilde M_{12}/\tilde M_{11}$ as functions of $\xi$ (we note that both $w_l$ and $w_t$ are independent of $q$). A more comprehensive study, including a comparison with some phenomenological formulas, can be found in~\cite{GSWP}.

In conclusion, we have shown that the short-range wakefields induced within the plasma bubble can be expressed in terms of matrix elements that can be calculated if the positions and momenta of the simulation macroparticles are known (recall that knowing the macroparticle coordinates is sufficient for determining the fields at a given $\xi$). Thus, apart from obtaining the details of the plasma flow, running our algorithm leads directly to useful information about the transverse and longitudinal wakefields.

\section{Conclusions}\label{sec:8}

In this paper, we have developed a novel, steady-state simulation technique that can deal with an axisymmetric PWFA configuration in the blowout regime. In particular, we have studied the propagation of two ultra-relativistic, non-evolving and axially symmetric bunches of arbitrary density profile through a cold plasma of uniform density. After formulating and analyzing the single-particle equations of motion for the plasma electrons, we show how the nonlinear plasma flow of the PWFA can be modeled using simulation macroparticles. In order to obtain a self-consistent description of the interaction, we combine
our analysis of the plasma dynamics with the equations that govern the formation of the electromagnetic fields. A crucial feature of our semi-analytical treatment is that, given the macroparticle coordinates at a given longitudinal position within the plasma bubble, we can determine the full pattern of the fields at that particular location. This decoupling between the transverse and longitudinal directions is what makes our approach different from the existing numerical algorithms~\cite{PhysRevSTAB.6.061301,doi:10.1063/1.872134,HUANG2006658} and allows us to determine the structure of the bubble in a systematic and efficient way. Using our technique, we have studied the beam loading effect for a Gaussian drive/witness beam configuration. Our results regarding the dimensions of the plasma cavity and the electric field pattern that is established within it are in agreement with those obtained from QuickPic, a well-known particle-in-cell PWFA code. In addition to this, our approach appears to offer a relative advantage in terms of computation speed, simplicity and versatility. The latter feature is particularly emphasized by the fact our algorithm can be used directly for the calculation of the short-range wakefields inside the plasma cavity.

\section{Acknowledgements}

We would like to thank V. Khudik for his important contributions to this work. We are thankful to X. Xu for providing the QuickPic data for the beam loading study and to W. An for the comparison of our algorithm with that of QuickPic. This work was supported by the Department of Energy, contract DE-AC03-76SF00515.
\appendix

%
\section{Dealing with delta functions and their derivatives in the equation for $\bar B_\theta$}\label{appendixA}
%

In this Appendix, we study the mathematical properties of a driven equation similar to the one for $\bar B_\theta$ (see Eqs.~\eqref{eq:19}, \eqref{eq:22} and \eqref{eq:23} in the main text). To start with, let us consider the following model equation for the function $y(x)$:
    \begin{align}\label{eq:A1}
    \frac{d^2y(x)}{dx^2}
    =
    ay(x)\delta(x-x_0)
    +
    b\delta(x-x_0)
    +
    c_0\delta'(x-x_0)
    ,
    \end{align}
where $x_0$, $a$, $b$ and $c_0$ are constants. Our goal is to obtain the matching conditions for $y$ and $y'$ at $x=x_0$ (note that both variables are discontinuous at the position of the singularity, on account of the presence of the delta function derivative on the RHS of Eq.~\eqref{eq:A1}). We first note that $\delta'(x-x_0) = h''(x-x_0)$, where $h(x-x_0)$ is the unit step function ($0$ for $x<x_0$ and $1$ for $x>x_0$) and re-write the above equation as
    \begin{align}\label{eq:A2}
    \frac{d^2[y(x)-c_0h(x-x_0)]}{dx^2}
    =
    [ay(x)+b]\delta(x-x_0)
    .
    \end{align}
Introducing $z(x) = y(x)-c_0h(x-x_0)$, we have
    \begin{align}\label{eq:A3}
    \frac{d^2z(x)}{dx^2}
    &=
    [az(x)+b]\delta(x-x_0)
    +
    ac_0h(x-x_0)\delta(x-x_0)
    =
    [az(x)+b]\delta(x-x_0)
    +
    \frac{1}{2}
    ac_0
    \frac{d}{dx}
    h^2(x-x_0)
    \nonumber\\
    &=
    \left[az(x)+b
    +
    \frac{1}{2}
    ac_0
    \right]\delta(x-x_0)
    ,
    \end{align}
where we have used the relations $\delta(x-x_0) = h'(x-x_0)$ and $h^2(x-x_0)=h(x-x_0)$. Integrating the above result from $x=x_0^{-}$ to $x=x_0^{+}$, we obtain
    \begin{align}\label{eq:A4}
    z'(x_0^+)-z'(x_0^-)
    =
    az(x_0)+b
    +
    \frac{1}{2}
    ac_0
    .
    \end{align}
We note that $z(x)$ is continuous at $x=x_0$ (the RHS of Eq.~\eqref{eq:A3}, as shown in the second line, does not contain a delta function derivative), so $z(x_0)=y(x_0^{-})$ is well defined. Collecting everything, we arrive at the matching conditions
    \begin{align}\label{eq:A5}
    y(x_0^{+})
    =
    y(x_0^{-})
    +
    c_0
    ,\qquad
    y'(x_0^{+})-y'(x_0^{-})
    =
    ay(x_0^{-})+b
    +
    \frac{1}{2}
    ac_0
    =
    \frac{1}{2}
    a
    [y(x_0^{+})+y(x_0^{-})]
    +
    b
    .
    \end{align}
When we have the operator $\p_x x^{-1}\p_x (xy)$ on the left-hand side of~\eqref{eq:A1}, we instead need to consider the equation
    \begin{align}\label{eq:A6}
    y''
    +
    \frac{1}{x}
    y'
    -
    \frac{y}{x^2}
    =
    ay(x)\delta(x-x_0)
    +
    b\delta(x-x_0)
    +
    c_0\delta'(x-x_0)
    .
    \end{align}
Eq.~\eqref{eq:A3} is now replaced by
    \begin{align}\label{eq:A7}
    z''
    +
    \frac{1}{x}
    [z'+c_0\delta(x-x_0)]
    -
    \frac{z + c_0h(x-x_0)}{x^2}
    =
    [az(x)+b]\delta(x-x_0)
    +
    \frac{1}{2}
    ac_0
    \frac{d}{dx}
    h^2(x-x_0)
    .
    \end{align}
Moving the term $c_0\delta(x-x_0)/x=c_0\delta(x-x_0)/x_0$ to the RHS, we have
 \begin{align}\label{eq:A8}
    z''
    +
    \frac{1}{x}
    z'
    -
    \frac{z + c_0h(x-x_0)}{x^2}
    =
    [az(x)+b-c_0/x_0]\delta(x-x_0)
    +
    \frac{1}{2}
    ac_0
    \frac{d}{dx}
    h^2(x-x_0)
    .
    \end{align}
Since $z$ is still continuous (no delta function derivatives among the driving terms), the added term proportional to $z'$ does not affect the matching conditions at $x=x_0$. The same is true of the term proportional to the combination $z+c_0h(x-x_0)$, which has a finite discontinuity at $x=x_0$. Thus, the sole effect of the change in the form of the LHS of~\eqref{eq:A1} is the replacement of $b$ with $b-c_0/x_0$ in~\eqref{eq:A5}.

By reviewing Eqs.~\eqref{eq:19}, \eqref{eq:22} and \eqref{eq:23}, we can easily verify that the equation for $\bar B_\theta$ is entirely analogous to the one studied above. In particular, the basic analogies are $x\rightarrow r$, $x_0\rightarrow r_i$ and $y\rightarrow \bar B_\theta$, so that we can use Eq.~\eqref{eq:A5} with
    \begin{align}\label{eq:A9}
    y(x_0^-)
    &\rightarrow
    a_{i-1}r_i
    +
    \frac{b_{i-1}}{r_i}
    ,\qquad
    y'(x_0^-)
    \rightarrow
    a_{i-1}
    -
    \frac{b_{i-1}}{r_i^2}
    ,
    \nonumber\\
    y(x_0^+)
    &\rightarrow
    a_{i}r_i
    +
    \frac{b_{i}}{r_i}
    ,\qquad
    y'(x_0^+)
    \rightarrow
    a_{i}
    -
    \frac{b_{i}}{r_i^2}
    \end{align}
and $a\rightarrow A_i$, $b\rightarrow B_i$ and $c_0\rightarrow C_i$. This leads directly to Eq.~\eqref{eq:25} in the main text.

\bibliography{\string~/gsfiles/Bibliography/master}%

\end{document}